%% file: main.tex
\documentclass[10pt,conference]{IEEEtran}
\usepackage{cite}
\usepackage{listings}
\usepackage{hyphenat}
\usepackage{amsmath,amssymb,amsfonts}
\usepackage{graphicx}
\usepackage{textcomp}
\usepackage{xcolor}
\usepackage{algorithm}
\usepackage{algpseudocode}
\usepackage{caption}
\usepackage{subcaption}
\usepackage{balance}
\usepackage{url}
\usepackage{comment}
\usepackage{booktabs} 

\renewcommand{\baselinestretch}{0.99}

\usepackage[normalem]{ulem} 
\usepackage{xcolor}
\newcommand{\raw}{$\rightarrow$}
\newcommand{\ugh}[1]{\textcolor{red}{\uwave{#1}}} 
\newcommand{\ins}[1]{\textcolor{blue}{\uline{#1}}} 
\newcommand{\del}[1]{\textcolor{red}{\sout{#1}}} 
\newcommand{\chg}[2]{\textcolor{red}{\sout{#1}}{\raw}\textcolor{blue}{\uline{#2}}} 

\usepackage{ifthen}
\newboolean{showcomments}
\setboolean{showcomments}{true} 
\ifthenelse{\boolean{showcomments}}
  {\newcommand{\nb}[2]{
    \fcolorbox{gray}{yellow}{\bfseries\sffamily\scriptsize#1}
    {\sf\small$\blacktriangleright$\textit{#2}$\blacktriangleleft$}
   }
   
  }
  {\newcommand{\nb}[2]{}
   
  }
\newcommand{\gena}[1]{\nb{Genaína}{\footnotesize #1}}

\newcommand{\ric}[1]{\nb{Ricardo}{\footnotesize #1}}
\newcommand{\eric}[1]{\nb{Eric}{\footnotesize #1}}

\AtBeginDocument{%
  \providecommand\BibTeX{{%
    \normalfont B\kern-0.5em{\scshape i\kern-0.25em b}\kern-0.8em\TeX}}}

\def\BibTeX{{\rm B\kern-.05em{\sc i\kern-.025em b}\kern-.08em
    T\kern-.1667em\lower.7ex\hbox{E}\kern-.125emX}}

\makeatletter
\newcommand{\linebreakand}{%
  \end{@IEEEauthorhalign}
  \hfill\mbox{}\par
  \mbox{}\hfill\begin{@IEEEauthorhalign}
}

\makeatletter
\def\ps@IEEEtitlepagestyle{
        \def\@oddfoot{\mycopyrightnotice}
        \def\@evenfoot{}
}
\def\mycopyrightnotice{
        {\footnotesize
                \begin{minipage}{\textwidth}
                        \centering
                        \textcopyright~2021 IEEE.  Personal use of this material is
permitted.  Permission from IEEE must be obtained for all other uses, in
any current or future media, including reprinting/republishing this
material for advertising or promotional purposes, creating new
collective works, for resale or redistribution to servers or lists, or
reuse of any copyrighted component of this work in other works.
                \end{minipage}
        }
}

\begin{document}

\makeatother

\title{Body Sensor Network: A Self-Adaptive System Exemplar in the Healthcare Domain}

\author{
\IEEEauthorblockN{Eric Bernd Gil}
\IEEEauthorblockA{
\textit{University of Bras\'ilia}\\
ericbgil@gmail.com}
\and
\IEEEauthorblockN{Ricardo Caldas}
\IEEEauthorblockA{
\textit{Chalmers \textbar \, University of Gothenburg}\\
ricardo.caldas@chalmers.se}
\and
\IEEEauthorblockN{Arthur Rodrigues}
\IEEEauthorblockA{
\textit{University of Bras\'ilia}\\
arthur.farias@aluno.unb.br}
\linebreakand
\IEEEauthorblockN{Gabriel Levi Gomes da Silva}
\IEEEauthorblockA{
\textit{University of Bras\'ilia}\\
gabrielevigomes@gmail.com}
\and
\IEEEauthorblockN{Gena\'ina Nunes Rodrigues}
\IEEEauthorblockA{
\textit{University of Bras\'ilia}\\
genaina@unb.br}
\and
\IEEEauthorblockN{Patrizio~Pelliccione}
\IEEEauthorblockA{
\textit{Chalmers \textbar \, University of Gothenburg}\\
\textit{Gran Sasso Science Institute (GSSI)}\\
patrizio.pelliccione@gssi.it}
}

\maketitle 

\input{abstract.tex}


\input{introduction.tex}
\input{background.tex}
\input{proposal.tex}
\input{experiments.tex}
\input{conclusion.tex}

\section*{Acknowledgment}

The authors express their utmost gratitude to Carlos Eduardo Taborda Lottermann and to Léo Moraes for supporting the implementation of the SA-BSN and to the members of the LADECIC research group for the fruitful discussions. 
This study was financed in part by CAPES-Brasil – Finance Code 001, through CAPES scholarship, by CNpq under grant number 306017/2018-0, by University of Brasilia under Call DPI/DPG 03/2020, by the Wallenberg Al, Autonomous Systems and Software Program (WASP) funded by the Knut and Alice Wallenberg Foundation. 

\balance
\bibliographystyle{IEEEtran}
\bibliography{ref}

\end{document}

%% file: abstract.tex
\begin{abstract}

Recent worldwide events shed light on the need of human-centered systems engineering in the healthcare domain. These 
systems must be prepared to evolve quickly but safely, according to unpredicted environments and ever-changing pathogens that spread ruthlessly. 
Such scenarios suffocate hospitals' infrastructure and disable healthcare systems that are not prepared to deal with unpredicted environments without costly re-engineering. 
In the face of these challenges, we offer the SA-BSN -- Self-Adaptive Body Sensor Network -- prototype to explore the rather dynamic patient's health status monitoring. The exemplar is focused on self-adaptation and comes with scenarios that hinder an interplay between system reliability and battery consumption that is available after each execution. Also, we provide: (i) a noise injection mechanism, (ii)  file-based patient profiles' configuration, (iii) six healthcare sensor simulations, and (iv) an extensible/reusable controller implementation for self-adaptation. The artifact is implemented in ROS (Robot Operating System), which embraces principles such as ease of use and relies on an active open source community support. 

\end{abstract}

%% file: introduction.tex
\section{Introduction}

In spite of various research efforts in IT for the healthcare domain in the last years, e.g.~\cite{abidi2017wireless,guk2019evolution}, 
the development of healthcare self-adaptive cyber-physical systems is still quite challenging due to its safety critical nature. The provision of assurance evidences for the compliance of quality attributes such as safety~\cite{reyes2016integration}, reliability~\cite{gravina2017multi}, and cost~\cite{sodhro2018energy} has a vital importance in this domain. Therefore,  equipping these systems with self-
configuration, self-healing, and self-management 
competences has become paramount for healthcare and assisted living applications.

The evolution of self-adaptive systems can be explained from a historical perspective through a set of complementary stages, namely \textit{waves}~\cite{weyns2020introduction}. 
These waves encompass concepts like: (i) automation of management tasks; (ii) architecture-based adaptation; (iii) use of models at runtime; (iv) adaptation based on goals; (v) provision of guarantees under uncertainties; (vi) control-based approaches; and (vii) use of learning techniques in adaptation. Although some of the concepts in \textit{vi} and \textit{vii} are well-established for at least five years~\cite{filieri2015software, esfahani2013learning}, and the inspiration from machine learning and control theory on self-adaptive systems goes way beyond that, we are just starting to grasp the benefits of adopting their fundamentals in the engineering of self-adaptive cyber-physical systems.

The SEAMS community has steadily supported in self-adaptive systems~\cite{seamsrepo}. Although there are proposals available in the healthcare domain~\cite{weyns2015tele}  and in the engineering of safety- and mission critical adaptive systems~\cite{wuttke2012traffic, gerasimou2017undersea, maia2019dragonfly}, to the best of our knowledge, there is no particular exemplar of a cyber-physical system based on control-theoretical principles in the domain of adaptive body sensor network systems. Driven by the need of having a control theory-based prototype to explore the effects of uncertainties on quality attributes of safety critical applications, we present the Self-Adaptive Body Sensor Network exemplar (SA-BSN)\cite{caldas_ricardo_2021_4620112}. The SA-BSN is designed as a software exemplar to monitor and analyse the health statuses of patients individually, through a set of sensors in tandem with a centralized processor. Moreover, as an exemplar for the self-adaptive domain, the adaptation goal of the SA-BSN is to keep a target reliability level while accounting for a target energy consumption management. In the past few years, our artifact has been not only under constant evolution but also adopted in the evaluation of some previous works of our research group~\cite{rodrigues2018learning,rodrigues2019enhancing,solano2019taming,Caldas2020hybrid}, proving itself as an artifact that is synergistic with the ideas presented in the aforementioned waves \textit{vi} and \textit{vii}. The use of our exemplar is straightforward, and we have made it available online, together with a virtual machine, supporting a set of scenarios and adaptation policies\footnote{\url{https://github.com/lesunb/bsn}}. The executable scenarios we propose demonstrate how our artifact adapts itself to cope with three distinct classes of uncertainty: the system itself, the system goals, and the environment.

The rest of the paper is structured as follows. In Section~\ref{sec:exemplar}, we present an overview of SA-BSN, provided with adaptation scenarios and the quality attributes involved. Section~\ref{sec:BSNdetails} describes the exemplar from the architecture and implementation perspectives, including implementation details as well. In Section~\ref{sec:experiment}, we guide the reader to use the provided artifact for experimentation. Finally, Section~\ref{sec:conclusion} concludes the paper with the next steps.

%% file: background.tex
\section{SA-BSN Exemplar and Adaptation Overview}\label{sec:exemplar}

The SA-BSN is an exemplar of a healthcare application implemented in ROS~\cite{quigley2009ros}. 
The goal of the SA-BSN is to detect emergencies by continuously monitoring the patient's health status. Furthermore, the SA-BSN is equipped to adapt itself in order to maintain the desired QoS levels with minimal human intervention, while accounting for classes of uncertainty. Hereafter we stand upon the well-known architectural view of managed and managing system as means of seeking for separation of concerns\cite{weyns2020introduction, braberman2015morph} to refer to the corresponding SA-BSN Managing and Managed System modules and their responsibilities. In this section, we describe the SA-BSN exemplar requirements in a goal-oriented perspective.

\begin{figure}[htbp]
    \centering
    \includegraphics[width=0.48\textwidth]{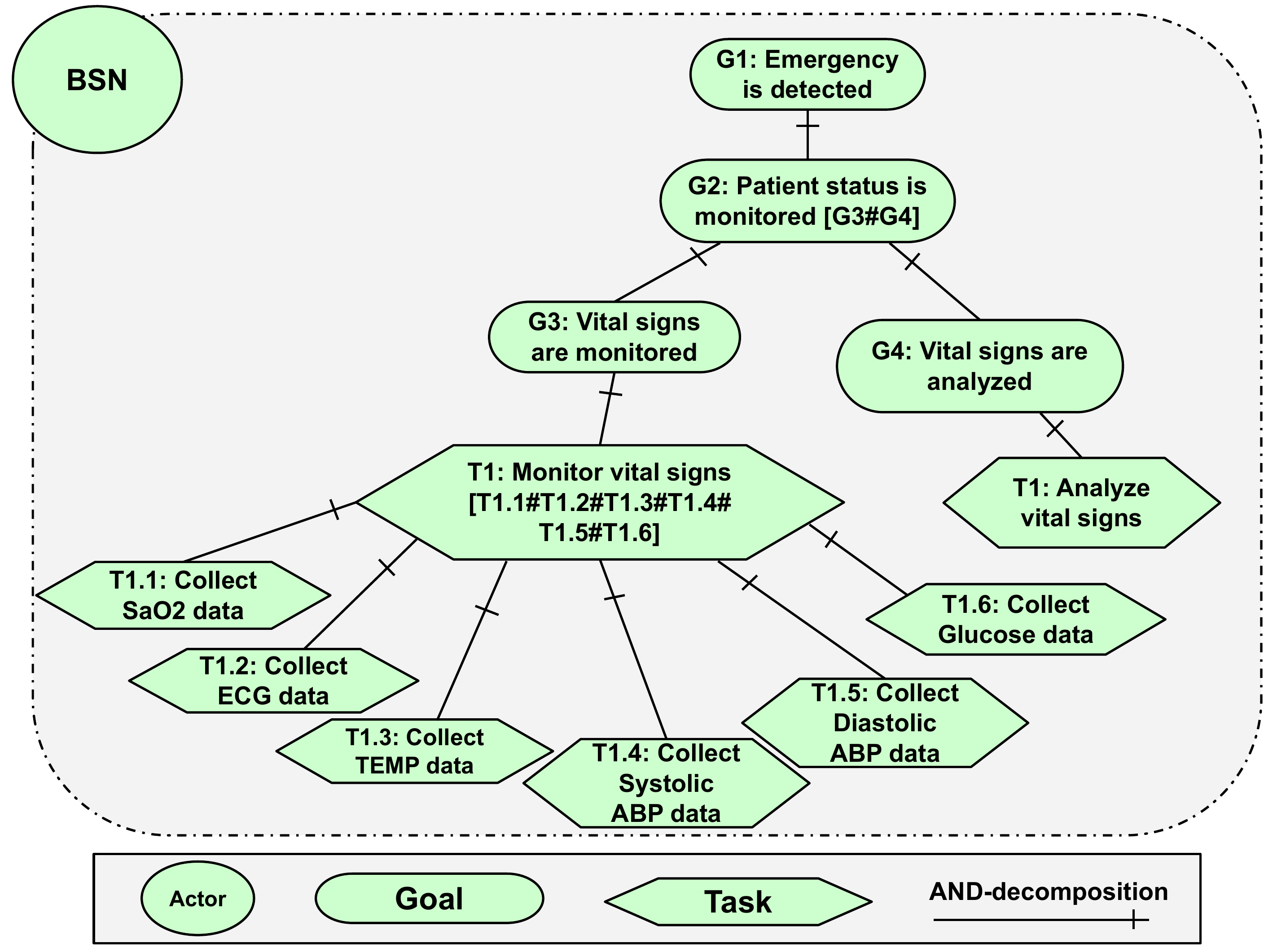}
    \caption{The Goal Model used to represent the BSN}
    \label{fig:BSN_GM}
\end{figure}

According to the goal-oriented view of the SA-BSN's functional requirements shown in Figure \ref{fig:BSN_GM}, the Managed System has six sensors to continuously monitor the patient's vital signs and achieve its goal of analysing the vital signs and detecting an emergency when it occurs. A range of vital signs is periodically collected from the patient through a set of distributed sensors: electrocardiograph sensor (ECG) for heart rate and electrocardiogram curve; a pulse oximeter (SaO2) for measuring blood oxygen saturation; a thermometer (TEMP), that collects the body temperature in Celsius; a sphygmomanometer for measuring  and systolic arterial blood pressure (ABP); there is also a Glucose sensor for measuring blood glucose levels. The collected data is then forwarded to the Central Hub: a component in the Managed System to fuse the vital signs and classify the overall health situation of the patient into \emph{low}, \emph{moderate}, or \emph{high} risk status. In addition to that, the Central Hub can stack other responsibilities like preprocessing the incoming data, filtering the redundancy, and translating communication protocols. 

The Managing System module of the SA-BSN is, in turn, responsible for continuously assuring the fulfillment of the desired QoS attributes related to the values of reliability and battery consumption (i.e. cost). For the evaluation of the quality of the adaptation we use control theory metrics following the terminology proposed by Camara et al.\cite{camara:2020:bridging}. The chosen QoS constraint is attributed to a setpoint, which is set by the user before the system execution. For example, if the concerned QoS attribute is the reliability, one could set it to 95\%, within an acceptable error range. This is called setpoint tracking, which can be measured by the steady-state error (SSE) metric. In addition to this requirement, the user can verify other control theoretic metrics related to the transient behavior, which can be evaluated at the end of the system execution. For instance, the user could set the threshold for the adaptation overshoot, specifying that it should not exceed 10\% of the target value (setpoint). Another example would be choosing the settling time in a way that the adaptation should not take more than 3 minutes to converge. We further illustrate these metrics in Section~\ref{sec:experiment}, where we present a running scenario for the BSN.

\begin{table*}[h]
\begin{tabular}{lllll}
\toprule
\textbf{Scenario} & \textbf{Uncertainty Class~\cite{solano2019taming} : Type}  & \textbf{Impact}                                                                                            & \textbf{Adaptation Policy}                                                                           & \textbf{Affected QoS Attribute} \\
\midrule
S1                & SI: unexpected number of users      & Delays on message processing                                                                               & Adjust Central Hub service time rate                                                                           & Reliability               \\
&   SG: uncertain sampling rate &   Uncertain mean time to failure  & Adjusting sensors' sampling rate & Reliability   \\\\
S2                & SG: uncertain sampling rate        & \begin{tabular}[c]{@{}l@{}}High battery consumption \end{tabular} & \begin{tabular}[c]{@{}l@{}}Adjusting sensors' sampling rate\end{tabular} & Cost       \\\\
S3                & ENV: uncertain sensor availability & Unwanted sensors activated                                                                          & Disable unwanted sensors                                                                    & Cost, reliability               \\
\bottomrule
\end{tabular}%
\caption{SA-BSN adaptation scenarios under different classes of uncertainty (SI: System Itself (S1); SG: System Goals (S1-S2); ENV: execution context (S3))}
\label{tab:scenarios}
\end{table*}




While trying to meet its requirements, the system is prone to a range of uncertainties. Thus, the controller is activated to mitigate the effects of unexpected events in quality attributes. Table~\ref{tab:scenarios} presents the uncertainties and the adaptation goals of three scenarios that can be executed in the SA-BSN. The first scenario, S1, focuses on uncertainties related to the overflow of sensed data into the the Central Hub queue and also to the possible data uncertainties in sensors, which are related to the reliability of the system. The second scenario, S2, focuses on the uncertainty related to the operational frequencies of the components, which can lead to a battery consumption that exceeds what is needed to satisfy the requirements. In the third scenario, S3, depending on the patient profile, the operator may not want to use certain sensors; with fewer components to manage, less uncertainty in the system is expected and, consequently, a more stable adaptation process.

%% file: proposal.tex
\section{SA-BSN Implementation Details} \label{sec:BSNdetails}

In this section we further delve into the architecture perspective (Section \ref{sec:architecture}) and the implementation details (Section \ref{sec:implementation}) of the SA-BSN exemplar.

\begin{figure}[htb]
    \centering
    \includegraphics[width=0.48\textwidth]{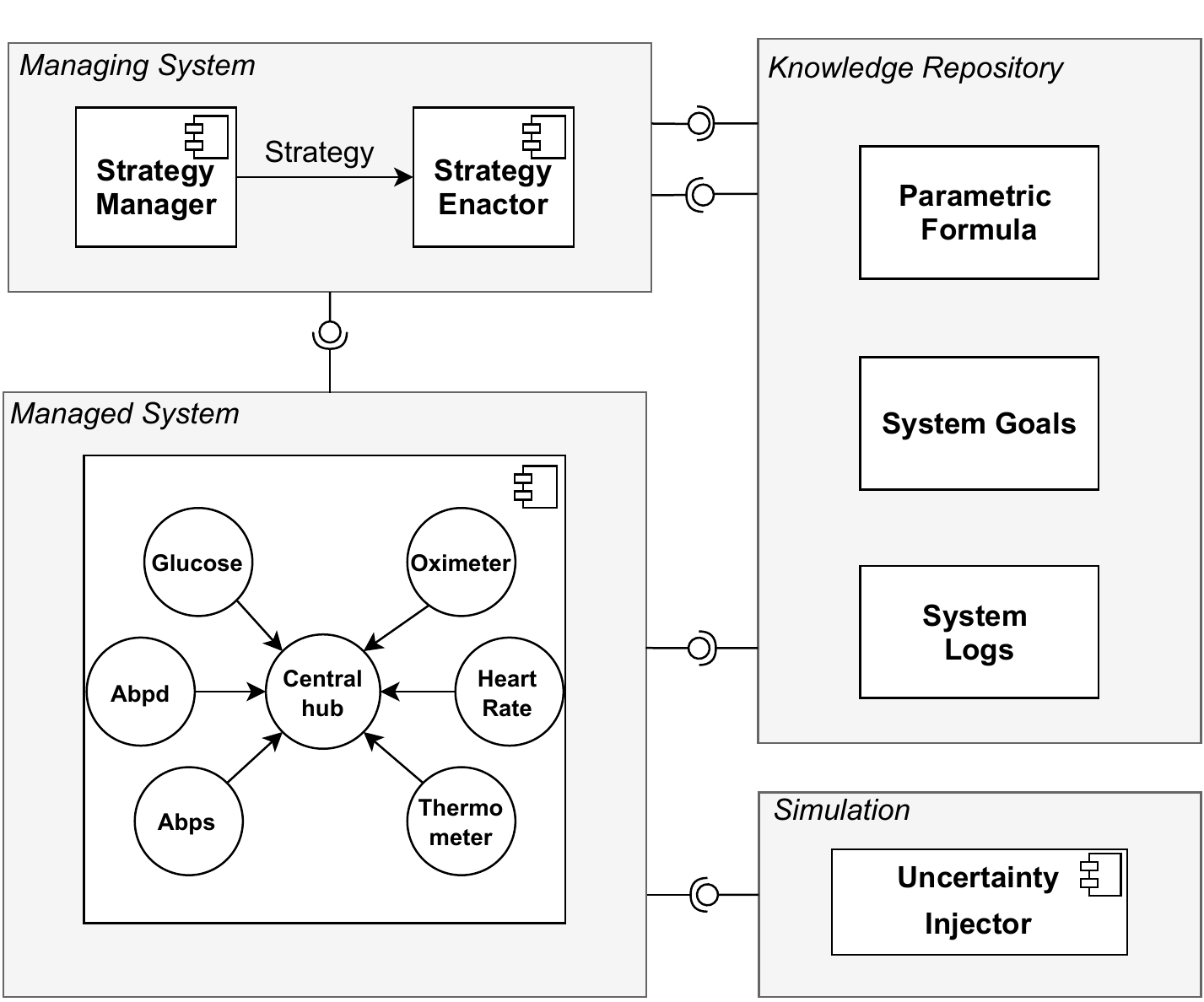}
    \caption{Architectural perspective of the SA-BSN as a self-adaptive exemplar.}
    \label{fig:BSNSAS-Arch}
\end{figure}

\subsection{SA-BSN architecture on ROS}\label{sec:architecture}

The SA-BSN artifact is composed of four main modules: Managing System, Managed System, Knowledge Repository and Simulation, as depicted in Figure~\ref{fig:BSNSAS-Arch}. Below we further detail these architecture modules of the SA-BSN and their functionalities.
\subsubsection{The Managing System}
This SA-BSN module comprises the Strategy Manager and Strategy Enactor and is in charge of implementing the controller to deal with the adaptation issues. The Strategy Manager is responsible for estimating the reliability and cost setpoints for the components of the Managed System module, given a system desired setpoint and the system's reliability and cost estimated via a parametric formula available in the knowledge repository \cite{Caldas2020hybrid}. The Strategy Enactor is where the controller is implemented; it is responsible for applying the adaptation strategies to achieve the previously estimated setpoints for each component. 

\begin{figure}[b]
    \centering
    \includegraphics[width=0.45\textwidth]{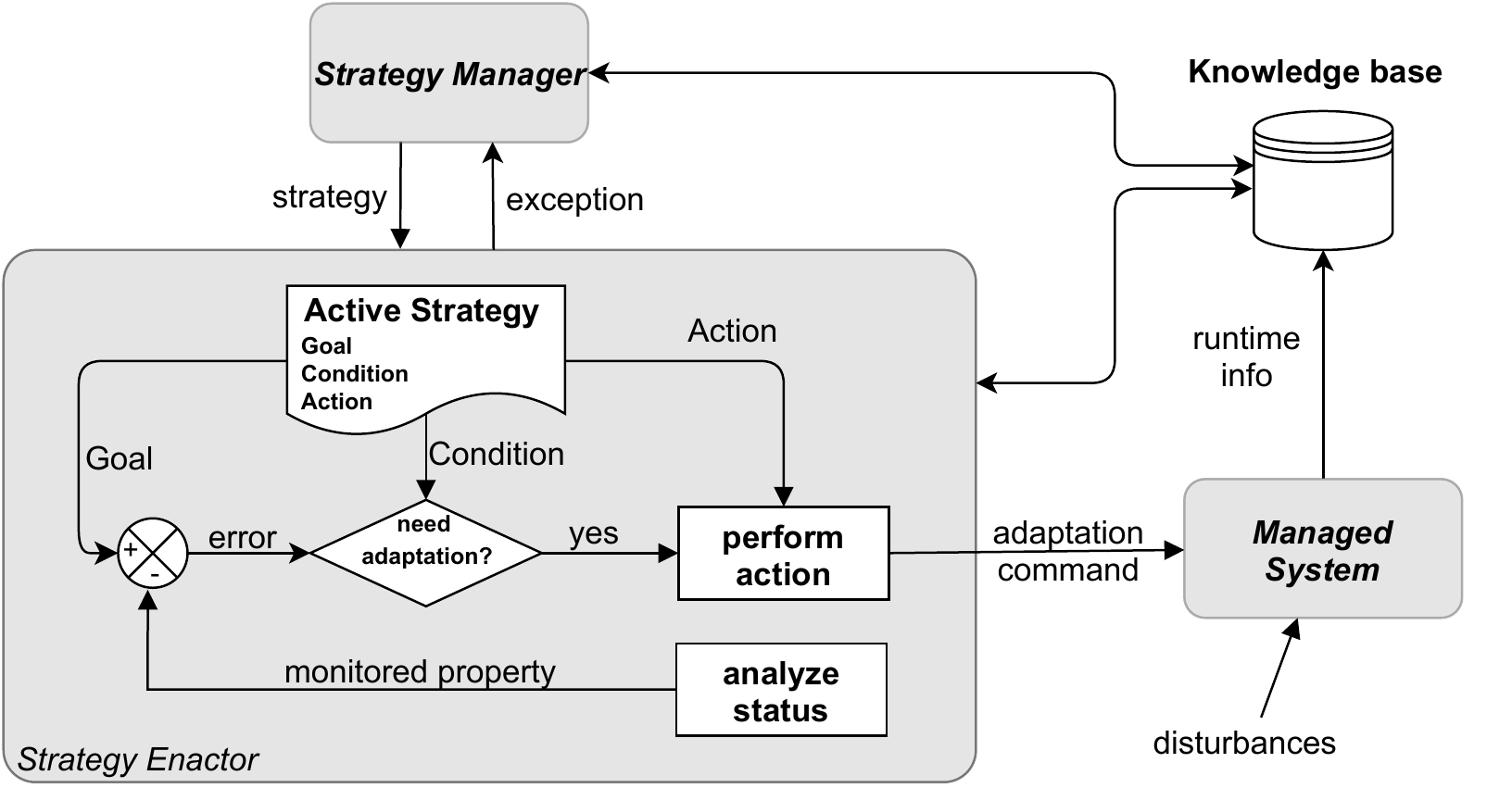}
    \caption{System feedback loop and its interactions~\cite{Caldas2020hybrid}}
    \label{fig:Enactor_Loop}
\end{figure}

Figure~\ref{fig:Enactor_Loop} shows interaction between these two components and the other parts of the system and 
the closed feedback loop for Strategy Enactor, which works for both the Managed System components' and system setpoints. This component uses the setpoints estimated by the Strategy Manager and compares them to the actual value of the desired QoS attribute (i.e., reliability or cost) for each component, applying the adaptation policy according to the analyzed status. In our case, the adaptation policy consists in adjusting the components frequencies according to the calculated error using the controller in order to simulate the sensors' sampling rate and the central hub processing rate. These knobs directly impact the overall system reliability since: (i) more data points collected by a sensor per time unit is expected to render a more precise measurement of the vital signal and (ii) messages lost in the Central Hub due to queue problems may impact the reliability of the system. Also, they impact the energy consumption (cost) of the components since more executions are performed per time unit. We should note that the BSN is not limited by actuation through such knobs or control based on reliability or cost. These are interchangeable with other actuation mechanisms and QoS attributes.


\subsubsection{The Managed System}

The Managed System comprises the sensors for vital signs monitoring and the Central Hub. The components responsible for the communication between this module and the other SA-BSN modules are the Probes and the Effectors. The Probes are responsible for gathering data from the Managed System components and sending them to the Knowledge Repository and the Managing System. The Effectors are responsible for receiving adaptation commands from the Strategy Enactor and changing the Managed System components parameters accordingly.

\subsubsection{The Knowledge Repository}
The knowledge repository comprises (i) the parametric formulas to compute the reliability and energy consumption of the Managed System, which were generated using our Pistar-GODA MDP artifact~\cite{solano2019taming}, (ii) the goals to be achieved, in the form of the goal model, and (iii) the System Logs where knowledge about the system's execution is persisted.The System Logs which consist of 5 different types of logs: Adaptation, Status, Event, Uncertainty and EnergyStatus. The Adaptation log is where the Strategy Enactor adaptation commands are persisted. The Status log is where information about Managed System components status is persisted. The Event log is where information about activation and deactivation of Managed System components is persisted. The Uncertainty log is where information about injected uncertainty is persisted. Finally, the EnergyStatus log is where cost information is persisted. We should note that the data persistence in the logs as well as the interface between the Knowledge Repository and the other SA-BSN modules is carried out by the Data Access component.

\subsubsection{The Simulation Module}
This fourth SA-BSN module comprises the Uncertainty Injector component to simulate the uncertainties envisioned for the Managed System. As such, this component is responsible for injecting uncertainty into the Managed System sensors in order to induce failure on the data collection process. One important comment to make here is that the sensors will not fail unless the uncertainty injector is active. However, if it is desirable that a specific group of sensors do not fail under any circumstances, one must configure the injector to not inject uncertainty into them; this feature is explained in the next sections.

\subsection{SA-BSN operation on ROS}\label{sec:implementation}

\begin{figure}[t]
    \centering
    \centerline{\includegraphics[width=0.5\textwidth]{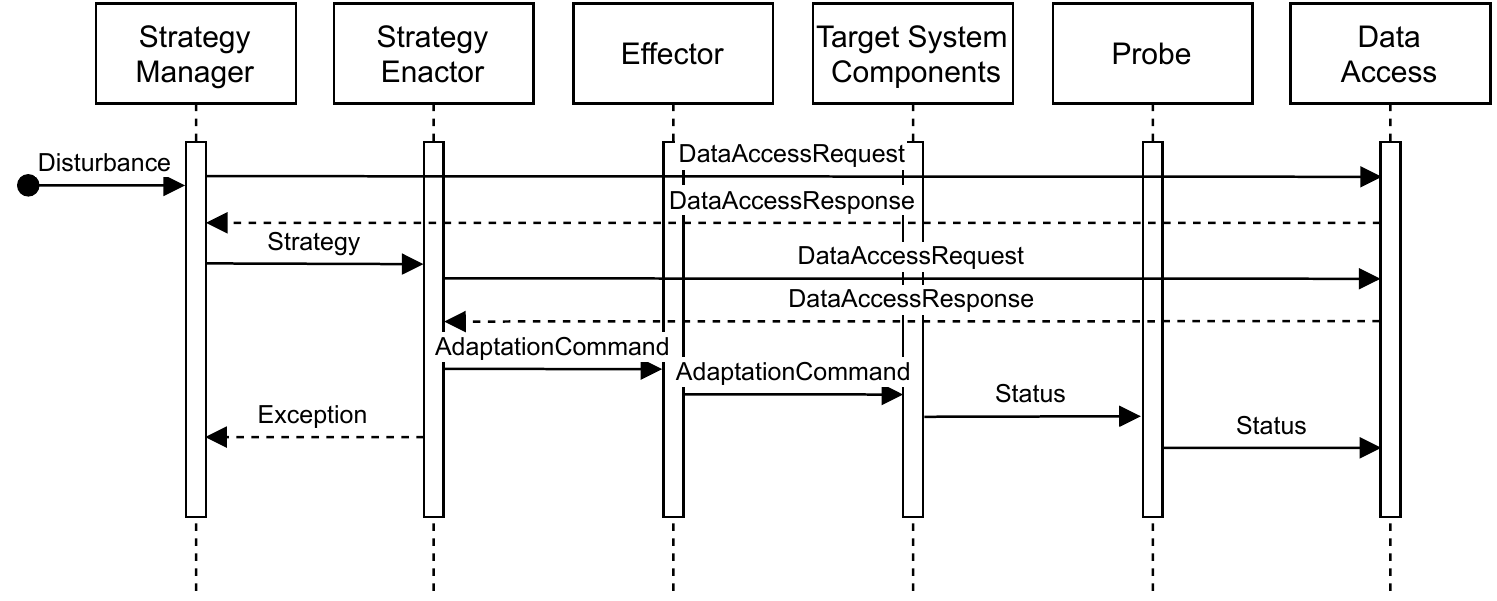}}
    \caption{Sequence diagram of a normal cycle of adaptation}
    \label{fig:SD_adaptation}
\end{figure}

The four SA-BSN modules are coordinated through ROS messages exchanged in a publish/subscribe architecture, using the TCP/IP communication protocol. The dynamic view of the adaptation process is shown in the sequence diagram in Figure \ref{fig:SD_adaptation}, where we present the messages exchanged between components when an adaptation is required. The need for adaptation is detected when there is a disturbance in the attribute of interest, i.e., when the error is bigger than the setpoint times a stability margin, which is fixed in 0.02. If the adaptation is needed, the Strategy Manager sends a \textit{DataAccessRequest} message to the Data Access component, which will fetch the log entries regarding failure rates or battery consumptions of the Managed System components. Then, the Strategy Manager estimates the setpoints for all Managed System components, and send them as a \textit{Strategy} message to the Strategy Enactor. In its turn, the Enactor uses these setpoints to estimate the frequency of the components, sending an \textit{AdaptationCommand} message to the Effector that redirects it to the target component. If any exception is verified, the Strategy Enactor sends an \textit{Exception} message to the Strategy Manager, which acts accordingly. Finally, the Managed System components receive the \textit{AdaptationCommand}, change their frequencies, and continue to send their current condition in periodic \textit{Status} messages to the Probe, which forwards them to the Data Access node right after;

%% file: experiments.tex
\section{Hands on the SA-BSN} \label{sec:experiment}

In this section, we present a guide to aid the reader in setting up an experimental environment and execute an adaptation scenario as a demonstration\footnote{Check our demonstration video in Youtube. Link provided in the github repository.} of the replicability of our exemplar.

\subsection{Customizing the Strategy Enactor}
We provide a default controller implementation into the Strategy Enactor component. We should note, however, that the researcher using the SA-BSN can implement its own controller. The objective of our default controller is to calculate frequency values for components according to the desired setpoint values for the adaptation metric calculated in the Strategy Manager. Furthermore, the user-defined controller builds the \textit{AdaptationCommand} message in order to communicate with the Effector. Similar to the other component in the system, the configuration of the controller and its parameters are defined in a launch file. Our default controller is a proportional controller, that consists in a proportional gain $K_{p}$, which acts directly on the error $e(t)$ between a desired setpoint and the current value of a given metric (e.g. reliability), generating a control output $u(t)$.
\begin{equation}
  u(t) = K_{p} \times e(t)
\end{equation}
For the case of a user-defined controller, the user must redesign the methods \textit{setUp}, \textit{apply\_reli\_strategy}, \textit{apply\_cost\_strategy}, and \textit{receiveEvent}, which are defined in the Controller source and header files. The \textit{setUp} method is responsible for reading the launch file and assigning values to the respective controller parameters as well as receiving the QoS attribute to be adapted, which is defined in the Strategy Manager. The \textit{apply\_reli\_strategy} and \textit{apply\_cost\_strategy} are the methods responsible for calculating the module's frequency values, depending on the QoS attribute used, building then the \textit{AdaptationCommand} message. Finally, the \textit{receiveEvent} method is responsible for receiving activation and deactivation signals, generated by the Managed System modules, and setting the parameters to default values. 

\subsection{System Configuration and Setup}

The setup required to build the runtime environment for the SA-BSN, comprises five stages: (i) experimental environment setup, (ii) vital signs generation setup (iii) sensors setup (iv) building the System Manager and (v) configuring the Uncertainty Injection mechanism. Below we further detail each stage. 

\subsubsection{Experimental Environment Setup}

There are two ways of obtaining the local version of the system, through the download of a virtual machine or the download of the source code. Further instructions to access both can be found in a Github repository\footnote{https://github.com/lesunb/bsn}, where we provide an installation guide.

In order to configure SA-BSN components parameters we use ROS features called launch files. They are XML-based files in which the tags are divided in three types: \verb+launch+, \verb+node+ and \verb+param+. The \verb+launch+ tag indicates the scope of the launch file. The \verb+node+ tag is responsible for setting up the component. The \verb+param+ tag\footnote{http://wiki.ros.org/roslaunch/XML/param} is responsible for the configuration of the components parameters and contains a \verb+name+ attribute, which represents the name of the parameter, a \verb+value+ attribute and an optional \verb+type+ attribute. 
When configuring the SA-BSN we mostly change the attribute value of the \verb+param+ tags.



\subsubsection{Vital Signs Generation Setup}
\begin{figure}[t]
  \includegraphics[width=\linewidth]{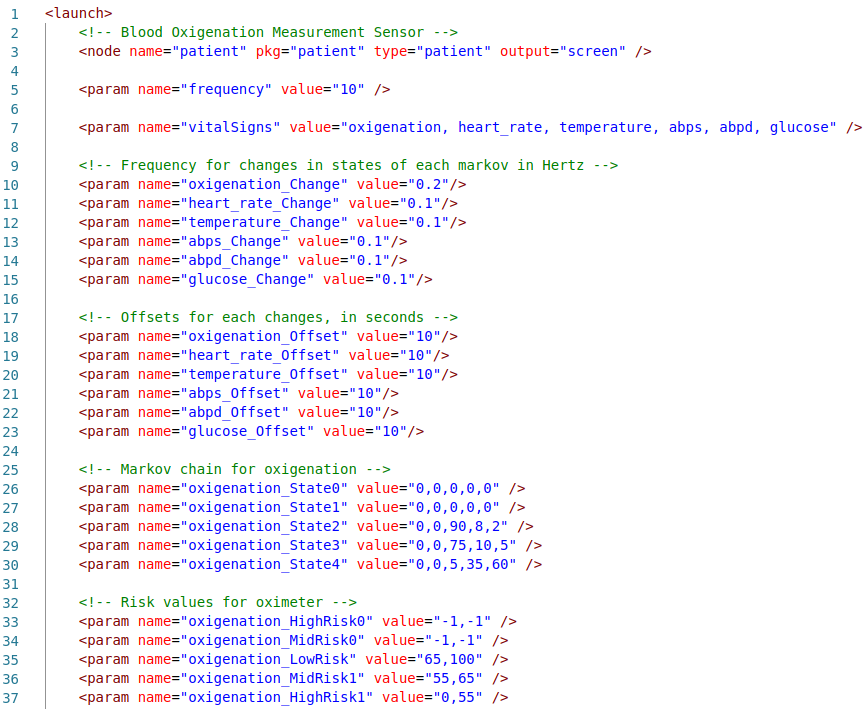}
  \caption{Patient module excerpt.} \label{fig:patient_launch_file}
\end{figure}

For the vital signs generation setup the user needs to configure the launch file for the Patient module, where an example of a partial configuration is shown in Figure \ref{fig:patient_launch_file}. 
This module is responsible for the generation of vital signs following a configured discrete-time markov chain model, so that sensors use ROS services to request for data in each execution. In the configuration file, we set the name of the vital signs to be generated (one for each sensor), as in line 7, the frequencies
for each change (change rate), as in lines 10-15, and the offset (in seconds), as in lines 18-23, where the data changes only if it has passed \textit{period} plus \textit{offset} seconds since the last change. We also define the transition probabilities for each of the five states the sensors can assume, as in lines 26-30,
and the range of risk values for each of these states, as in lines 33-37.

\subsubsection{Sensors setup}


 \begin{figure}[b]  \includegraphics[width=0.8\linewidth]{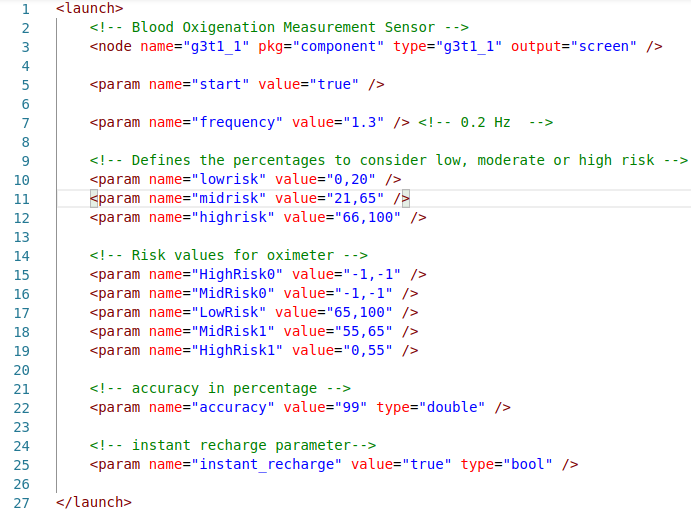}
  \caption{Sensor module excerpt} \label{fig:sensor_launch_file}
 \end{figure}%

For the sensors setup, the user needs to configure one launch file for each sensor. See Figure \ref{fig:sensor_launch_file}, for an example of the Oximeter's configuration. As shown in lines 10-12, the probability occurrences are configured for low, moderate and high risks. In lines 15-19, the range of values are defined for the vital signs risks. The third and fourth parameters are the sensor accuracy, defined in line 22, and the instant\_recharge parameter, defined in line 25. The instant\_recharge specifies whether the sensor simulates the recharging of the battery or not. Finally, the last parameter is the start parameter, shown in line 5, which defines whether the sensor will be active during the execution or if it will be shutdown right at the beginning.


\subsubsection{Building the System Manager}

The System Manager components configuration entail the Strategy Enactor and the System Manager launch files. 

To configure the Strategy Enactor, the user has to configure the frequency and $K_{p}$, since we provided  a proportional controller as default. In case of a user-defined controller, we could have as many parameters as needed. 

There are several parameters to be defined to configure the Strategy Manager: \textit{monitor\_freq}, \textit{setpoint}, \textit{actuation\_freq}, \textit{info\_quant}, \textit{offset}, \textit{gain}, and \textit{qos\_attribute}. The \textit{monitor\_freq} is the frequency in which the Strategy Manager will monitor the values of the chosen adaptation metric. The \textit{actuation\_freq} is the frequency in which the Strategy Manager will calculate values for the setpoints of the Managed System components. The \textit{setpoint} parameter is the system setpoint to be achieved, where the system reliability and cost are estimated via their corresponding parametric formula. The \textit{info\_quant} parameter specifies how many data points will be used to calculate the reliability.
The \textit{offset} and \textit{gain} parameters are related to the search algorithm, together they delimit the size of the adaptation space. Finally, the \textit{qos\_attribute} parameter is name of the QoS attribute of interest. In order to use the engine for adaptation of reliability one must have to set the node parameter name and type attributes to "reli\_engine" while for cost one must have to set them to "cost\_engine".



\begin{figure}[t]
\includegraphics[width=\linewidth]{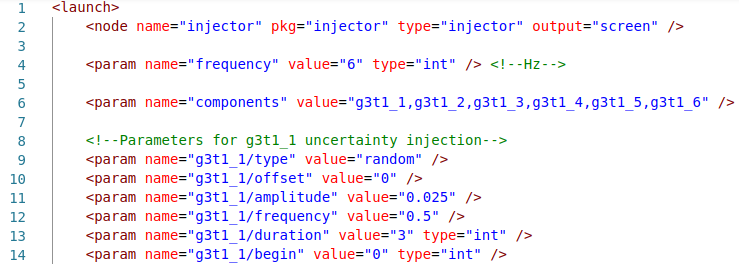}
\caption{Uncertainty injector excerpt} \label{fig:injector_launch_file}
\end{figure}

\subsubsection{Configuring the Uncertainty Injection Mechanism}
\begin{figure}[b]
  \centering
  \centerline{\includegraphics[width=0.5\textwidth]{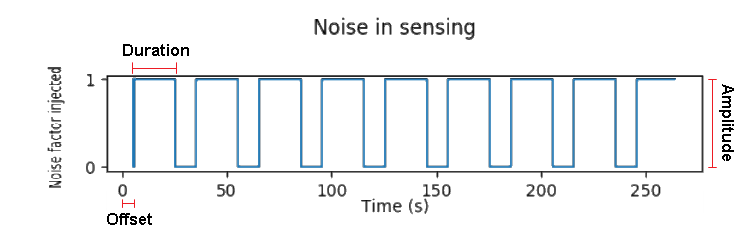}}
  \caption{Example of a step uncertainty signal waveform}
  \label{fig:wave_forms}
\end{figure}

\begin{figure*}[t]
  \centering
  \centerline{\includegraphics[width=0.85\textwidth]{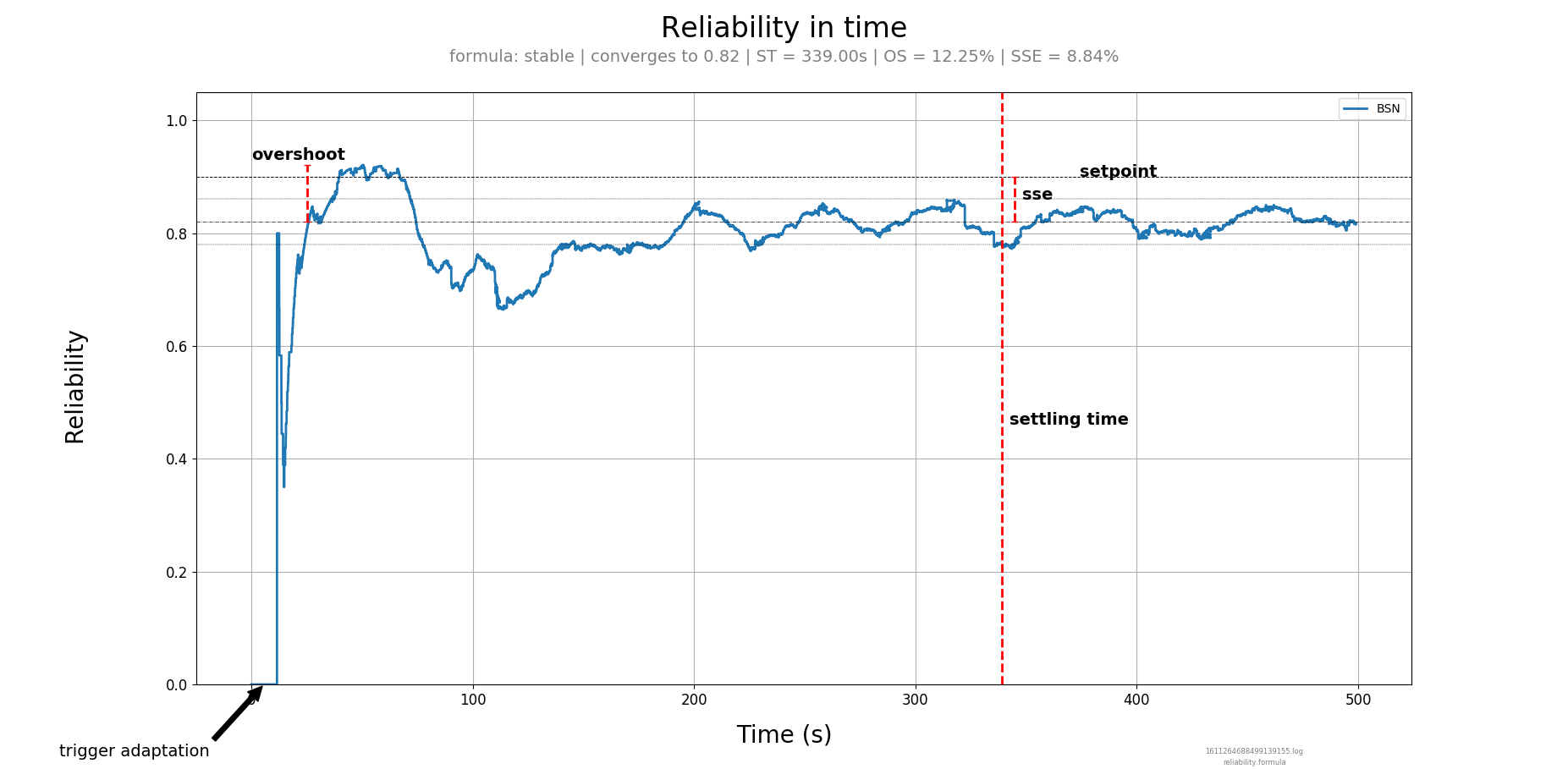}}
  \caption{Reliability curve obtained after the execution}
  \label{fig:rel_curve}
\end{figure*}

The last setup stage required to run the SA-BSN is the configuration of the uncertainty injection mechanism, where noise is introduced into the sensors to induce failure by data inconsistency. An example of a partial configuration for this module is shown in Figure~\ref{fig:injector_launch_file}. The configurable parameters for this module are the injection frequency (line 4) and the sensors in which uncertainty shall be injected (line~6).




In lines 9-14, we refer to the configuration of a few other parameters that can be also considered for each sensor: type, offset, amplitude, frequency, duration, and begin. The \textit{type} parameter defines the kind of wave the uncertainty will follow, which can be either \verb+step+, \verb+ramp+ or \verb+random+. An example of a \verb+step+ waveform is shown in Figure~\ref{fig:wave_forms}, where the wave parameters set in the configuration are indicated. It is important to notice that each component can be stimulated by a different type of noise. The \verb+offset+ parameter refers to the uncertainty offset to be injected. The \verb+amplitude+ parameter is the magnitude of the uncertainty value, which has implications on how much the noisy data will differ from the original collected data. The value of such a difference is what defines if the sensor failed or not due to the noisy input. The \verb+frequency+ parameter is the rate in which the uncertainty must be injected in the Managed System sensors. It is worth mentioning that, for sensors with instantaneous recharge (instant\_recharge parameter), the \verb+frequency+ parameter defines an upper bound of the number of failures that can happen.

\subsection{Example Scenario Evaluation}



The user must execute the ready-to-go bash script (\textit{run.sh}) to run the exemplar.
The run command can be configured by maximum execution time (seconds) as an argument (e.g. ``bash run.sh 30"). The default duration of the execution is 300 seconds, in case that no argument is passed.

Once the SA-BSN runs, a series of terminals will pop up on the screen, each corresponding to a component. By these means, the user can keep track of the execution progress.
After the execution time has elapsed, the terminals will close, and the user can check the logs or run the \textit{analyzer.py} script (inside the ``src/simulation/analyzer" folder) to obtain a graph describing the monitored QoS attribute curve, i.e., reliability or cost. Including the aforementioned control-theoretic metrics of interest. Instructions for the use of the \textit{analyzer.py} script are given in the SA-BSN GitHub repository. 

To illustrate our exemplar, we exercise scenario S1~(cf. Table~\ref{tab:scenarios}) and plot the results in Figure \ref{fig:rel_curve}. In scenario S1, we simulate the SA-BSNs reliability degrading due to processing message delays: an unexpected number of patients are simultaneously using the system, flooding the communication channel of the Central Hub. The analyzer.py script generates Figure \ref{fig:rel_curve} that contains 540 seconds of execution with all the six sensors active, which was obtained by the artifact's default configuration. Furthermore, we can verify that the setpoint, in this case, was 90\% reliability and the convergence value was 82\%, which resulted in a steady-state error (SSE) of 8.84\%. The maximum value reached was 112.25\% of the convergence value that results in an overshoot of 12.25\%. Finally, the settling time was 339 seconds. 

%% file: conclusion.tex
\section{Conclusion} \label{sec:conclusion}

In this paper we provide  SA-BSN, an exemplar of the self-adaptive system in the healthcare domain. Our exemplar sheds light on control theoretical solutions for SAS by providing an environment with disturbance injection in the components that encode vital signs monitoring sensors and a central hub. Henceforth, we present the requirements that guided the SA-BSN implementation and the uncertainty scenarios available in the current version. Then, we discuss the implementation details in ROS and provide a walk-through for interested users.
